\documentclass{desyproc}

\input{paperdef}
\usepackage{amsmath,amssymb,color,cite}
\graphicspath{{figs/}}

\begin{document}

\thispagestyle{empty}
\setcounter{page}{0}
\def\thefootnote{\fnsymbol{footnote}}

\begin{flushright}
\mbox{}
DCPT/10/152\\
DESY 10-152 \\
IPPP/10/76 
\end{flushright}

\vspace{1cm}

\begin{center}

{\large\sc {\bf BSM Higgs Physics at the LHC in the Forward Proton Mode}}
\footnote{talk given by S.H.\ at the {\em Physics at the LHC 2010}, 
May 2010, DESY Hamburg, Germany}

\vspace{1cm}

{\sc 
S.~Heinemeyer$^{1}$%
\footnote{
email: Sven.Heinemeyer@cern.ch}%
, V.A.~Khoze$^{2,3}$%
\footnote{
email: V.A.Khoze@durham.ac.uk}%
, M.G.~Ryskin$^{2,4}$%
\footnote{
email: Ryskin@MR11084.spb.edu}%
,\\[.5em] M.~Tasevsky$^{5}$%
\footnote{
email: Marek.Tasevsky@cern.ch}%
~and G.~Weiglein$^{6}$%
\footnote{email: Georg.Weiglein@desy.de}
}

\vspace*{1cm}

{\it
$^1$Instituto de F\'isica de Cantabria (CSIC-UC), 
Santander,  Spain\\

\vspace{0.3cm}
$^2$IPPP, Department of Physics, Durham University, 
Durham DH1 3LE, U.K.\\

\vspace{0.3cm}
$^3$School of Physics \& Astronomy, University of Manchester, 
Manchester M13 9PL, U.K.\\

\vspace{0.3cm}
$^4$Petersburg Nuclear Physics Institute, Gatchina, 
St.~Petersburg, 188300, Russia\\

\vspace{0.3cm}
$^5$Institute of Physics, 
18221 Prague 8, Czech Republic

\vspace{0.3cm}

$^6$DESY, Notkestra\ss e 85, D--22607 Hamburg, Germany
}
\end{center}

\vspace*{0.2cm}

\BC {\bf Abstract} \EC
We review the prospects for Central Exclusive Diffractive (CED)
production of Higgs bosons in the SM with a fourth generation of
fermions at the LHC using 
forward proton detectors installed
at 220~m and 420~m distance around ATLAS and / or CMS.
We discuss the determination of Higgs spin-parity and coupling
structures at the LHC and show that the forward proton mode
would provide a critical information on the $\cp$ properties
of the Higgs bosons.   

\def\thefootnote{\arabic{footnote}}
\setcounter{footnote}{0}

\newpage


\title{BSM Higgs Physics at the LHC in the Forward Proton Mode}


%

%
\author{{\slshape Sven Heinemeyer$^1$\footnote{Speaker}, 
                  V.A.~Khoze$^{2,3}$, 
                  M.G.~Ryskin$^{2,4}$,
                  M.~Tasevsky$^5$,
                  Georg Weiglein$^6$}\\[1ex]
$^1$Instituto de F\'isica de Cantabria (CSIC-UC), Santander, Spain\\
$^2$IPPP, Department of Physics, Durham University, 
Durham DH1 3LE, U.K.\\
$^3$School of Physics \& Astronomy, University of Manchester, 
Manchester M13 9PL, U.K.\\
$^4$Petersburg Nuclear Physics Institute, Gatchina, 
St.~Petersburg, 188300, Russia\\
$^5$Institute of Physics, 
18221 Prague 8, Czech Republic\\
$^6$DESY, Notketra{\ss}e 85, 22607 Hamburg, Germany}

%

\contribID{xy}  
\confID{1964}
\desyproc{DESY-PROC-2010-01}
\acronym{PLHC2010}
\doi            

\maketitle

\begin{abstract}
We review the prospects for Central Exclusive Diffractive (CED)
production of Higgs bosons in the SM with a fourth generation of
fermions at the LHC using 
forward proton detectors installed
at 220~m and 420~m distance around ATLAS and / or CMS.
We discuss the determination of Higgs spin-parity and coupling
structures at the LHC and show that the forward proton mode
would provide a critical information on the $\cp$ properties
of the Higgs bosons.
\end{abstract}


\section{Introduction}

In the recent years there has been a growing interest 
in the possibility to complement the standard LHC searches a Higgs boson
by the options offered by forward and diffraction physics. 
These assume the installation of
near-beam proton detectors in the LHC tunnel
installed at 220~m and 420~m  around ATLAS and / or CMS, see
\citeres{ar,KMRProsp,acf,DKMOR,KMRbsm,fp420rev} and references therein.
The combined detection of the centrally produced system and
both outgoing protons can 
provide valuable information on the Higgs sector of MSSM and other popular
BSM scenarios~\cite{KKMRext,diffH,CLP,acf,tripl}.
Another simple example of physics beyond the SM is a
model which extends the SM by a fourth generation of heavy fermions
(SM4), see, for instance,~\cite{extra-gen-review,4G,four-gen-and-Higgs}. 
Here it is assumed that the masses of the 4th generation quarks
are (much) heavier than the mass of the top-quark.
In this case, the effective coupling of the Higgs boson
to two gluons is three times larger than in the SM, and all branching
ratios change correspondingly.

The central exclusive diffractive (CED) processes are of the form
\begin{align}
pp\to p \oplus H \oplus p~, 
\end{align}
where the $\oplus$ signs denote
large rapidity gaps on either side of the   centrally produced state.
However, proving that a detected new state is, indeed, a Higgs boson 
will be far from trivial.
In particular, it will be of great importance 
to determine the spin and
$\cp$ properties of a new state and to measure precisely its mass, width
and couplings.

\newcommand{\sixoo}{60 \ifb}
\newcommand{\sixooeff}{60 \ifb\,eff$\times2$}
\newcommand{\sixooo}{600 \ifb}
\newcommand{\sixoooeff}{600 \ifb\,eff$\times2$}

\newcommand{\TOT}{{\rm tot}}
\newcommand{\SMv}{{\rm SM4}}
\newcommand{\HSM}{H^{\SM}}
\newcommand{\HSMv}{H^{\SMv}}
\newcommand{\MHSM}{M_{\HSM}}
\newcommand{\MHSMv}{M_{\HSMv}}

Following  \cite{diffH} we
consider four luminosity
scenarios: ``\sixoo'' and ``\sixooo'' refer to running at low and high 
instantaneous luminosity, respectively, using conservative assumptions
for the signal rates and the experimental
sensitivities; possible improvements of both theory and experiment 
 could allow for the scenarios where the
event rates are higher by a factor of 2, denoted as ``\sixooeff'' and
``\sixoooeff''.


\section{The Higgs boson in the SM4}

A simple example of physics beyond the SM is a
model, ``SM4'', which extends the SM by a fourth generation of heavy
fermions, see, for instance, \citeres{extra-gen-review,4G,4G-ew}. 
In particular, the masses of the 4th generation quarks
are assumed to be (much) heavier than the mass of the top-quark
(whereas the masses of the 4th generation leptons, which do not play a role
  here, are less restricted). 
In this case, the effective coupling of the Higgs boson
to two gluons is three times larger than in the SM.
No other coupling, relevant to LEP and Tevatron searches, changes
significantly.
Essentially, only the partial decay width
$\Gamma(H\to gg)$ changes by a factor of 9 and, with it, the 
total Higgs width and therefore all the decay branching ratios, see for
instance \citere{four-gen-and-Higgs,ggH4}.
The new total decay width and the relevant decay branching ratios can be
evaluated as,\\[-1em]
\begin{align}
\Gamma_\SM(H\to gg) &= \br_\SM(H\to gg)\:\Gamma_\TOT^\SM(H)\,,\\ 
\Gamma_\SMv(H\to gg) &= 9\:\Gamma_\SM(H\to gg)\,,\\
\Gamma_\TOT^\SMv(H) &= \Gamma_\TOT^\SM(H) - \Gamma_\SM(H\to gg)
	+ \Gamma_\SMv(H\to gg)\,.
\end{align}
%
The Higgs boson searches at LEP~\cite{LEPHiggsSM,LEPHiggsMSSM}
have been re-interpreted by {\tt HiggsBounds}~\cite{higgsbounds} 
in the SM4. 
The bound on the SM Higgs boson at LEP of $\MHSM \ge 114.4 \gev$ at the
95\%~C.L. is modified to $\MHSMv \ge 112 \gev$.
On the other hand Higgs boson searches in the SM4 at the
Tevatron~\cite{SM4-CDF-D0} have been performed. 
The range $130 \gev \lsim \MHSMv \lsim 210 \gev$ is found to be
excluded. Combining the two analyses leaves us with a window of allowed
Higgs masses in the SM4 of $112 \gev \lsim \MHSMv \lsim 130 \gev$.

\begin{figure*}[htb!]
\vspace{-1em}
\begin{center}
\includegraphics[width=0.49\textwidth]{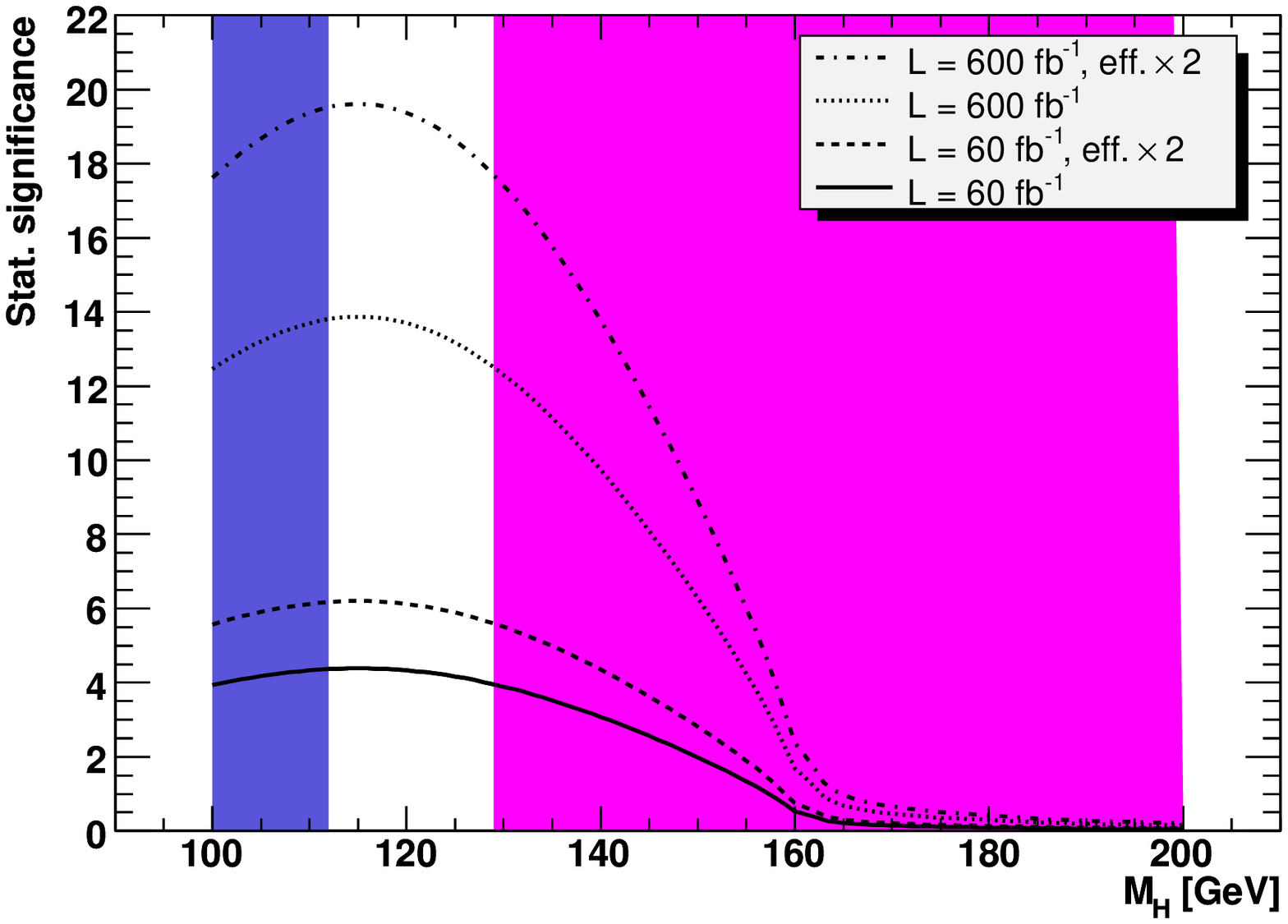}\;
\includegraphics[width=0.49\textwidth]{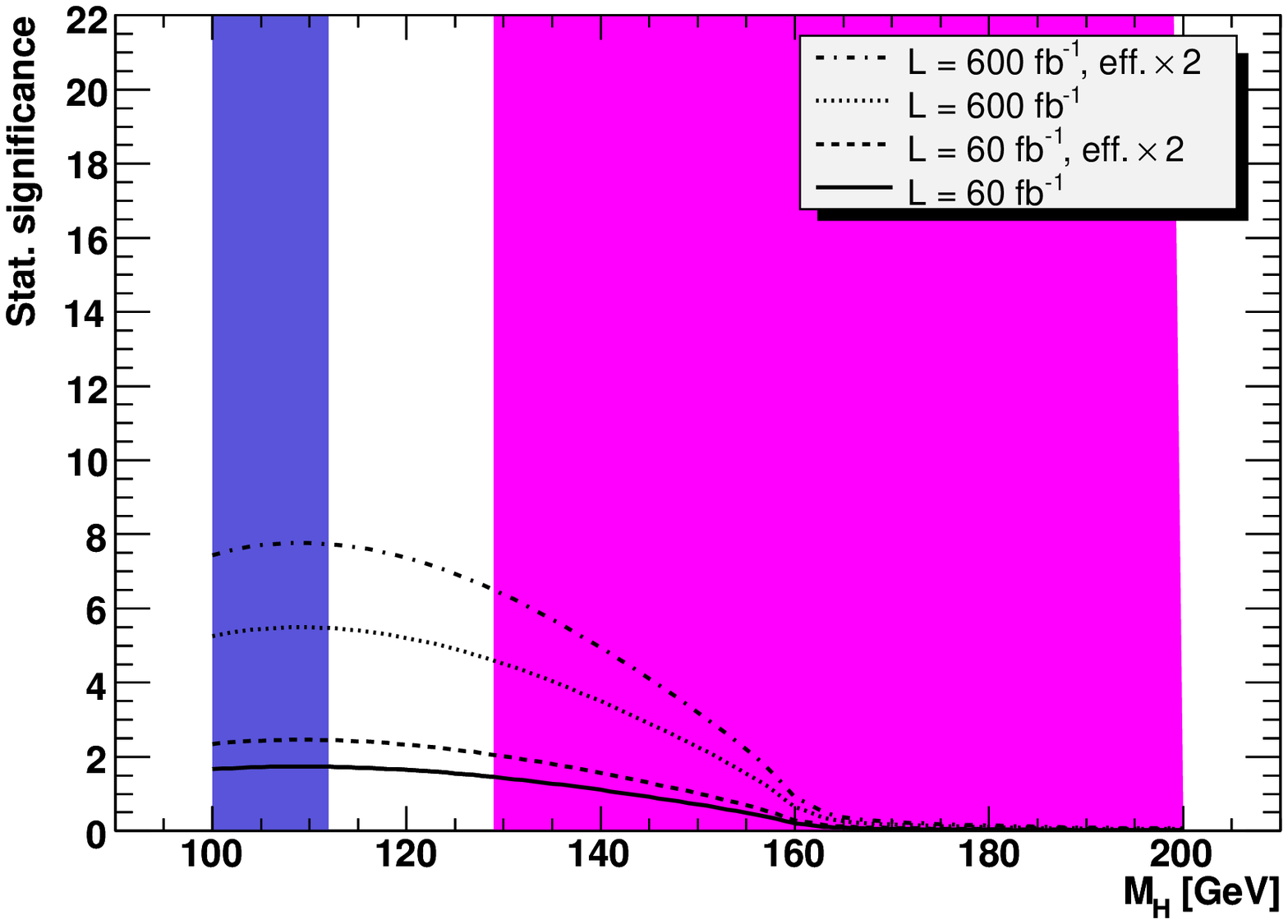}
\end{center}
\vspace{-1.5em}
\caption {
Significances reachable CED Higgs production in the SM4
  in the $H \to b \bar b$ 
  (left plot) and $H \to \tau^+\tau^-$ (right plot) channel 
for effective luminosities of ``\sixoo'',
``\sixooeff'', ``\sixooo'' and ``\sixoooeff''.
The regions excluded by LEP appear as blue/light grey for low values of 
$\MHSMv$ and excluded by the Tevatron as red/dark grey for larger values of
$\MHSMv$.} 
\label{fig:disc-SM4}
\end{figure*}

We have evaluated the significances that can be obtained
in the channels $H \to b \bar b$ and $H \to \tau^+\tau^-$. The results
are shown in \reffi{fig:disc-SM4} as a function of $\MHSMv$ for the four
luminosity scenarios. 
The regions excluded by LEP appear as blue/light grey for low values of 
$\MHSMv$ and regions excluded by the Tevatron appears as red/dark grey
for larger values of $\MHSMv$. 
The $b \bar b$ channel (left plot) shows that even at rather low
luminosity the remaining window of $112 \gev \lsim \MHSMv \lsim 130 \gev$
can be covered by CED Higgs production. Due to the smallness of 
$\br(\HSMv \to b \bar b)$ at $\MHSMv \gsim 160 \gev$, however, the CED
channel becomes irrelevant for the still allowed high values of
$\MHSMv$, and we do not extend our analysis beyond 
$\MHSMv \le 200 \gev$. The $\tau^+\tau^-$ channel (right plot) has not enough
sensitivity at low
luminosity, but might become feasible at high LHC luminosity.
At masses $\MHSMv \gsim 220 \gev$ it might be possible to exploit the decay 
$H \to WW, ZZ$, but no detailed analysis has been performed up to now.


\section{Coupling structure and spin-parity determination}

The determination of the spin and the $\cp$ properties of Higgs
bosons using the standard methods rely to a large extent on the coupling of a
relatively heavy SM-like Higgs to two gauge bosons.
The first channel that should be mentioned here is $H \to ZZ \to 4l$.
This channel provides detailed information about spin and $\cp$-properties
if it is open~\cite{jakobs-rev}.
Within a SM-like set-up it was analyzed how the tensor structure of the 
coupling of the Higgs boson to weak gauge bosons can be determined at the 
LHC~\cite{HVV-LHC0,HVV-LHC2,HVV-LHC1}.
One study for $\MHSM = 160 \gev$ was based on Higgs production in weak
vector boson fusion with the subsequent decay to SM gauge boson. 
It was shown that the discrimination between the two extreme
scenarios of a pure $\cp$-even (as in the SM) and a pure $\cp$-odd
tensor structure at a level of 4.5 to  5.3\,$\si$ using only about 10~\ifb.
A discriminating power of two standard deviations at $\MHSM = 120 \gev$
in the tau lepton decay mode requires an integrated luminosity of
30~\ifb~\cite{HVV-LHC1}. 

For $\MH \approx \MA \gsim 2 \MW$ the lightest MSSM
Higgs boson couples to gauge bosons with about SM strength, but its mass
is bounded from above by $\Mh \lsim 135 \gev$~\cite{mhiggsAEC},
i.e.\ the light Higgs stays below the threshold above which the decay to
$WW^{(*)}$ or $ZZ^{(*)}$ can be exploited.
On the other hand, the 
heavy MSSM Higgs bosons, $H$ and $A$, decouple from the gauge bosons.
Consequently, the analysis for $\MHSM = 160 \gev$ cannot be taken over
to the MSSM.
This shows the importance of channels to determine spin and
$\cp$-properties of the Higgs bosons without relying on (tree-level)
couplings of the Higgs bosons to gauge bosons.
CED Higgs production can yield crucial information in this
context~\cite{KMRProsp,KKMRext,diffH}.
 
The $\MHSM = 120 \gev$ analysis, on the other hand, 
can in principle be applied to the SUSY case. However, in this case the
coupling of 
the SUSY Higgs bosons to tau leptons, does not exhibit a
(sufficiently) strong enhancement as compared to the SM
case. Consequently, no improvement over the $2 \si$~effect within the SM
can be expected. 
The same would be true in any other model of new physics
with a light SM-like Higgs and heavy Higgses that decouple from
the gauge bosons.

The CED
production channels may provide crucial information on the $\cp$
properties of Higgs-like states detected at the LHC, for instance via
the $J_z$ selection rule~\cite{KMRmm}.
Thanks to this selection rule in the CED case we already know
that the observed new object has even parity (${\cal P} = +$) 
and the projection of its spin is $J_z=0$. 
This knoweledge will greatly simplify the determination of the
spin of new boson.

As discussed in \cite{KKMRext,diffH} it will be challenging to identify
a $\cp$-odd Higgs boson, for instance the $A$~boson of the MSSM, in the
CED processes. 
The strong suppression, caused by the
$P$-even selection rule, effectively filters out its production.
However in the semi-inclusive diffractive reactions the pseudoscalar
production is much less suppressed. As shown in a recent study~\cite{KMRbsm} 
there are certain advantages of looking for the $\cp$-odd Higgs particle
in the semi-inclusive process $pp \to p + gAg + p$
with  two tagged forward  protons and two large rapidity gaps.
The amplitude of $\cp$-odd $A$~boson
production can be of the same order as the $\cp$-even boson amplitude if 
events with relatively hard gluons, whose energy
is comparable with the energy of the whole $g A g$ system, are selected.


\section{Acknowledgments}

Work supported in part by the European Community's Marie-Curie Research
Training Network under contract MRTN-CT-2006-035505
`Tools and Precision Calculations for Physics Discoveries at Colliders'
(HEPTOOLS).




\begin{footnotesize}

\newcommand{\ea}{{\it et al.}}

\end{footnotesize}



\begin{thebibliography}{99}

%
%

\bibitem{ar} M.~Albrow and A.~Rostovtsev,
             arXiv:hep-ph/0009336.

\bibitem{KMRProsp} V.A.~Khoze, A.D.~Martin and M.~Ryskin,
                   {\em Eur. Phys. J.} {\bf C 23} (2002) 311.

\bibitem{acf} M.~Albrow, T.~Coughlin and J.~Forshaw,
              arXiv:1006.1289 [hep-ph].

\bibitem{DKMOR} A.~De~Roeck, V.A.~Khoze, A.~Martin, R.~Orava and M.~Ryskin, 
                {\em Eur. Phys. J.} {\bf C 25} (2002) 391.

\bibitem{KMRbsm} V.~A.~Khoze, A.~Martin, M.~Ryskin and A.~Shuvaev,
                 {\em Eur.\ Phys.\ J.} {\bf C 68} (2010) 125.

\bibitem{fp420rev} M.~Albrow et al.\ [FP420  R\&D Collaboration],
                  {\em JINST} {\bf 4} (2009) T10001.

\bibitem{KKMRext} A.~Kaidalov, V.A.~Khoze, A.D.~Martin and M.~Ryskin, 
                  {\em Eur. Phys. J.} {\bf C 33} (2004) 261.

\bibitem{diffH} S.~Heinemeyer, V.~A.~Khoze, M.~G.~Ryskin, 
                W.~J.~Stirling, M.~Tasevsky and G.~Weiglein,
                {\em Eur.\ Phys.\ J.} {\bf C 53} (2008) 231.

\bibitem{CLP} B.~Cox, F.~Loebinger and A.~Pilkington, 
              {\em JHEP} {\bf 0710} (2007) 090.

\bibitem{tripl} M.~Chaichian, P.~Hoyer, K.~Huitu, V.~A.~Khoze and
                A.~Pilkington,
                {\em JHEP} {\bf 0905} (2009) 011.

\bibitem{extra-gen-review} P.~Frampton, P.~Hung and M.~Sher,
                           {\em Phys.\ Rept.} {\bf 330} (2000) 263.

\bibitem{4G} B.~Holdom et al., 
             {\em PMC Phys.} {\bf A 3} (2009) 4 
             [arXiv:0904.4698 [hep-ph]].

\bibitem{four-gen-and-Higgs} G.~Kribs, T.~Plehn, M.~Spannowsky and T.~Tait,
                             {\em Phys.\ Rev.} {\bf D 76} (2007) 075016.

\bibitem{4G-ew} J.~Erler and P.~Langacker,
                {\em Phys.\ Rev.\ Lett.} {\bf 105} (2010) 031801.

\bibitem{ggH4} C.~Anastasiou, R.~Boughezal and E.~Furlan,
               {\em JHEP} {\bf 1006} (2010) 101.

\bibitem{LEPHiggsSM} G.~Abbiendi et al.\  [ALEPH, DELPHI, L3, OPAL
                     Collaborations and LEP Working Group for Higgs
                     boson searches], 
                     {\em Phys.\ Lett.} {\bf B 565} (2003) 61.

\bibitem{LEPHiggsMSSM} S.~Schael et al.  [ALEPH, DELPHI, L3, OPAL
                       Collaborations and LEP Working Group for Higgs
                       boson searches], 
                       {\em Eur.\ Phys.\ J.} {\bf C 47} (2006) 547.

\bibitem{higgsbounds} P.~Bechtle, O.~Brein, S.~Heinemeyer, G.~Weiglein
                      and K.~Williams, 
                      {\em Comput.\ Phys.\ Commun.} {\bf 181} (2010) 138;
                      arXiv:0909.4664 [hep-ph];
                      see: {\tt www.ippp.dur.ac.uk/HiggsBounds}~.

\bibitem{SM4-CDF-D0}   TEVNPH Working Group Collaboration 
  for the CDF Collaboration and D0 Collaboration,
  CDF note 10101, D\O\ note 6039.

\bibitem{jakobs-rev} V.~Buescher and K.~Jakobs,
                     {\em Int.\ J.\ Mod.\ Phys.} {\bf A 20} (2005) 2523.

\bibitem{HVV-LHC0} T.~Plehn, D.~Rainwater and D.~Zeppenfeld,
                   {\em Phys.\ Rev.\ Lett.}  {\bf 88} (2002) 051801.

\bibitem{HVV-LHC2} V.~Hankele, G.~Klamke, D.~Zeppenfeld and T.~Figy,
                   {\em Phys.\ Rev.} {\bf D 74} (2006) 095001.

\bibitem{HVV-LHC1} C.~Ruwiedel, N.~Wermes and M.~Schumacher,
                   {\em Eur.\ Phys.\ J.} {\bf C 51} (2007) 385.

\bibitem{mhiggsAEC} G.~Degrassi, S.~Heinemeyer, W.~Hollik,
                    P.~Slavich and G.~Weiglein, 
                    {\em Eur. Phys. J.} {\bf C 28} (2003) 133.

\bibitem{KMRmm} V.A.~Khoze, A.~Martin and M.~Ryskin, 
                {\em Eur. Phys. J.} {\bf C 19} (2001) 477
                [Erratum-ibid.\ {\bf C 20} (2001) 599].


\end{thebibliography}
\end{document}